\documentclass[final,3p,times,twocolumn]{elsarticle}
\usepackage{lineno,hyperref}
\modulolinenumbers[5]

\makeatletter
\def\ps@pprintTitle{%
  \let\@oddhead\@empty
  \let\@evenhead\@empty
  \let\@oddfoot\@empty
  \let\@evenfoot\@oddfoot
}
\makeatother

\usepackage{subfigure}
\usepackage[ruled,vlined]{algorithm2e}

\begin{document}

\begin{frontmatter}

\title{Class Token and Knowledge Distillation for Multi-head Self-Attention Speaker Verification Systems}

\author{Victoria Mingote, Antonio Miguel, Alfonso Ortega, Eduardo Lleida}
\address{ViVoLab, Arag\'{o}n Institute for Engineering Research (I3A), University of Zaragoza, Spain.}
\address{\{vmingote,amiguel,ortega,lleida\}@unizar.es}




\begin{abstract}
This paper explores three novel approaches to improve the performance of speaker verification (SV) systems based on deep neural networks (DNN) using Multi-head Self-Attention (MSA) mechanisms and memory layers.
Firstly, we propose the use of a learnable vector called Class token to replace the average global pooling mechanism to extract the embeddings.
Unlike global average pooling, our proposal takes into account the temporal structure of the input what is relevant for the text-dependent SV task.
The class token is concatenated to the input before the first MSA layer, and its state at the output is used to predict the classes.
To gain additional robustness, we introduce two approaches.
First, we have developed a new sampling estimation of the class token.
In this approach, the class token is obtained by sampling from a list of several trainable vectors.
This strategy introduces uncertainty that helps to generalize better compared to a single initialization as it is shown in the experiments.
Second, we have added a distilled representation token for training a teacher-student pair of networks using the Knowledge Distillation (KD) philosophy, which is combined with the class token.
This distillation token is trained to mimic the predictions from the teacher network, while the class token replicates the true label.
All the strategies have been tested on the RSR2015-Part II and DeepMine-Part 1 databases for text-dependent SV, providing competitive results compared to the same architecture using the average pooling mechanism to extract average embeddings.
\end{abstract}


\end{frontmatter}

\nolinenumbers

\section{Introduction}
The performance in speaker verification (SV) tasks has improved greatly in recent years thanks to the deep learning (DL) advances in signal representations and optimization metrics \cite{Taigman2014DeepFace:Verification,snyder2018x,Hoffer2015DeepNetwork,schroff2015facenet,snyder2016deep} that have been adapted from the state-of-the-art face verification, image recognition, or text-modelling systems.
In these systems, Convolutional Neural Network (CNN) or Time Delay Neural Network (TDNN) \cite{snyder2018x} are still the most employed approaches to obtain the signal representations or embeddings.
Nevertheless, self-attention mechanisms are becoming a dominant approach in many fields beyond text-related tasks.
For example, Transformers \cite{vaswani2017attention} are spreading to many tasks \cite{kenton2019bert,dosovitskiy2020image,touvron2020training,locatello2020object} where large scale databases are available.
In SV tasks, this kind of architecture has started to be successfully applied in text-independent SV \cite{india2019self,shim2022graph,wang2022multi,han2022local} where there are no constraints in the uttered phrase and big databases are available.
However, in text-dependent SV, there is still room for improvement since the amount of public data is not very large.
Besides, text-dependent SV consists of deciding whether a speech sample has been uttered by the correct speaker pronouncing the fixed passphrase selected.
So, the phonetic information of the signal is relevant to determine the identity.
Therefore, keeping the temporal structure is needed to obtain representations that encode correctly both phrase and speaker information.

%
In the context of text-dependent SV tasks, our previous works \cite{Mingote2018,mingote2019supervector,mingote2020optimization} showed the advantages of replacing the traditional pooling mechanism based on averaging the temporal information with an external alignment mechanism to obtain a supervector embedding.
This supervector allowed to keep the temporal structure and represent both phrase and speaker information, but the temporal alignment had to be performed by using an external method as a phone decoder, a Gaussian Mixture Model (GMM) \cite{reynolds1995robust,reynolds2000speaker} or a Hidden Markov Model (HMM) \cite{rabiner1989tutorial}.
As an alternative approach, in \cite{Mingote2021icassp}, we introduced Multi-head Self-Attention (MSA) mechanisms \cite{vaswani2017attention} combined with memory layers \cite{lample2019large} to substitute the alignment mechanisms.
The use of MSA allowed the model to focus on the most relevant frames of the sequence to discriminate better among utterances and speakers.
However, the proposed architecture based on MSA employed an average pooling mechanism to obtain the final representation embedding.

In this work, to substitute the global average pooling, we have introduced a learnable vector known as Class token, which is inherited from Natural Language Processing (NLP) \cite{kenton2019bert}, and recently, many image recognition systems \cite{dosovitskiy2020image}.
However, this approach has not yet been applied to SV tasks.
To introduce this vector into the system based on DNN with MSA and memory layers, the class token is concatenated to the input before the first MSA layer, and the state at the output is employed to perform the class prediction.
During training, the temporal information is encoded in the token, and this token interacts with the whole input sequence through self-attention and learns a global description similar to a supervector approach \cite{mingote2019supervector,cai2018novel} since the multiple heads act as slots of the supervector.
A similar mechanism has also been used recently in \cite{locatello2020object}.
Therefore, the average pooling mechanism is not needed to obtain a representation.
The multiple heads can encode more details about the sequence order than the average, playing the role of the states and improving the results as shown in \cite{mingote2019supervector},\cite{mingote2020optimization} with the use of external alignment mechanisms based on HMM and GMM.
In addition, the information encoded in these multiple heads can be represented and analyzed, which improves the interpretability of the results of this kind of approach.
%
To improve the performance obtained with the class token approach, we also introduce a novel multiple initialization sampling mechanism to reduce possible initialization problems and give more robustness against the lack of data to model predictions.
%
Since it is a case of use in the industry to develop custom specific systems with small in-domain datasets and this kind of approach could be a possible solution.

Moreover, this work contributes with another approach based on Transformer architecture and Knowledge Distillation (KD) \cite{Hinton2015DistillingNetworkb,touvron2020training}.
We propose a teacher-student approach combined with Random Erasing data augmentation \cite{zhong2020random,mingote2020bdk} which allows modelling the uncertainty in the parameters of a teacher model with a compact student model and get more reliable predictions.
Following the idea proposed in \cite{touvron2020training}, we have also introduced the Distillation token in the student network to replicate the predictions of the teacher network, while the class token is trained to reproduce the true label as Fig.\ref{fig3} depicts.  
%
Unlike the objective in \cite{touvron2020training}, in our work, the distillation process is not intended to compress the teacher model, but rather both models are trained together and the student model learn to better capture the intrinsic variability of the teacher predictions.

To summarize, the main contributions are:
\begin{itemize}
    \item We replace the global average pooling mechanism by a learnable class token to obtain a global utterance descriptor associated to the concept of supervector in speaker verification.
    \item We propose a new approach based on a sampling approximation to estimate the class token.
    \item We introduce a teacher-student architecture with an additional token known as distillation token which is combined with the class token to provide robustness to the learned student model.
\end{itemize}

This paper is organized as follows. 
In Section 2, we show an overview of the MSA and memory layers. 
Section 3 explains the strategy of introducing a learnable class token using sampling. 
In Section 4, we introduce the approach based on KD combined with the tokens employed to develop our system.
Section 5 describes the system used.
In Section 6, we present the experimental data, and Section 7 explains the results achieved. 
Conclusions are presented in Section 8.

\section{Overview of Transformer Encoder}
The original transformer architecture \cite{vaswani2017attention} is composed of two main parts: the encoder and decoder parts.
However, in many tasks, the transformer encoder is the only part used to create the DL systems.
The core mechanism of each encoder block is Multi-head Self-Attention (MSA) layer which is composed of multiple dot-product attention.
As we only employ the encoder part, the input to this attention mechanism is the same for the query, key and value signals ($Q, K, V$):
\begin{equation}
Q_{h}= x \cdot {W_{h}^Q}, \ K_{h}= x \cdot {W_{h}^K}, \ V_{h}= x \cdot {W_{h}^V},
\end{equation}
where $x$ is the input to this layer, and $W_{h}^Q, W_{h}^K, W_{h}^V$ are learnable weight matrices to make the linear projections.
After these projections, a softmax operation is performed over the temporal axis, which allows each head to focus on certain frames of the input sequence.
The result of this softmax operation is known as the self-attention matrix for each head and can be defined as: 
\begin{equation}
A_{h}= softmax_{t}\left(\frac{Q_{h} \cdot K_{h}^\intercal}{\sqrt d_{k}}\right) ,
\label{eqah}
\end{equation}
where $d_{k}$ is the number of dimensions of the query/key vector, and $^{\intercal}$ denotes transpose.
This self-attention matrix learns the most relevant information among the different data.
Using this information, the value $V$ feature vectors are aggregated to obtain the output of each head. 
The final output of each head can be calculated as,
\begin{equation}
H_{h}= A_{h} \cdot {V_{h}}.
\label{eqh}
\end{equation}
%
Thus, MSA is defined as the concatenation of the outputs from each head $H_{h}$:
\begin{equation}
MSA (X)= [H_{1}, H_{2} \dots H_{d^{head}}] \cdot {W^{head}},
\end{equation}
where $X$ is the input to the attention layer, $W^{head}$ is a learnable weight matrix to make a final linear projection, and $d^{head}$ is the number of attention heads in the $h-th$ layer.

The transformer encoder alternates the MSA layer with a second layer which is the feed-forward (FF) layer.
However, in \cite{Mingote2021icassp}, we proposed the replacement of FF layers by memory layers as in \cite{lample2019large}.
With this layer, the input data is compared with all the keys using a product key-attention, and the scores obtained are used to select the closest keys, which have the highest scores.
After that, the associated weight vectors are computed with the following expression:
\begin{equation}
w=softmax_{n}(x \cdot U^K),
\label{eq1}
\end{equation}
where $x$ is the input to the layer, $U^K$ is the keys matrix, and the softmax is computed over the memory index axis to focus on certain contents of the memory that will be used to provide the output.
Once these vectors are obtained, these weights are combined with the memory values of the selected keys, and the output is concatenated with the previous attention output:
\begin{equation}
x_{out}=x+w\cdot U^V,
\end{equation}
where $w$ are the weights of the selected keys obtained with (\ref{eq1}), and $U^V$ are the memory values associated with the keys.
After the encoder blocks are applied, an average pooling mechanism is usually employed to reduce the temporal information and represent variable-length utterances with fixed-length vectors.
%
However, this averaging may neglect the order of the phonetic information, which is relevant for text-dependent SV tasks.

\section{Representation using Class Token}

In many tasks of NLP and computer vision, the transformer architecture uses a learnable vector called Class Token ($x_{CLS}$), as in the original BERT model \cite{kenton2019bert} or Vision Transformer (ViT) \cite{dosovitskiy2020image}, instead of a global average pooling. 
To employ this token in the transformer encoder, the vector is concatenated to the input of the first MSA layer to perform the classification task.
With this token, the self-attention is forced to capture the most relevant information with the class token to obtain a representation as a global utterance descriptor similar to the supervector approach.
%
%
Instead of mixing all the information with an average pooling mechanism, the temporal structure can be kept since the attention mechanism acts as a weighted sum of the temporal tokens for each layer.
%
The output vector is the concatenation of different head subvectors and each of them is the result of a different attention outcome.
Thus, the mechanism can be seen similar to those used in our previous work \cite{mingote2019supervector}, where the heads play the role of the states and the supervector in \cite{cai2018novel}.
The supervector mechanism is also similar to \cite{vinals2019phonetically} but in that case, the task was text-independent SV and MSA layers were not used.
Besides, this type of mechanism allows to enhance the interpretability of what the neural network learns through the self-attention layers.

In \cite{cai2018novel}, this mechanism to obtain the supervector is defined similar to a conventional GMM supervector with the following expression:

\begin{equation}
s_{c}=\frac{\sum_{t}x_{t}\cdot w_{tc}}{\sum_{t}w_{tc}}=\sum_{t}x_{t}\cdot \bar{w}_{tc},
\label{eqsv}
\end{equation}
where $w_{tc}$ are the weights obtained by a softmax function on the output of a learnable layer, $s_{c}$ are vectors per state/component \textit{C} of dimension $D$ that summarize the information associated along the sequence of feature vectors $x_{t}$ of dimension $D$, and $\bar{w}_{tc}$ are the normalized weights defined as $ w_{tc}/{\sum_{t}w_{tc}}$.
The final supervector is built by the concatenation of these vectors $S=\{s_{1}, ... , s_{C}\}$ and is used to represent the whole sequence.
In this work, the output feature vectors for each head $H$ of the MSA layer are obtained with (\ref{eqh}) as a weighted sum equivalent to (\ref{eqsv}), where $\bar{w}_{tc}$ corresponds to the rows of the matrix of self-attention weights $A_{h}$ obtained with (\ref{eqah}).
%
In particular, for the class token, the normalized weights would be obtained from the last row of $A_{h}$.
Therefore, the final class token obtained with this mechanism is the concatenation of the different head subvectors corresponding to the class token position, which can be expressed as the supervector presented previously $S_{CLS}=\{s_{1-CLS}, ... , s_{H-CLS}\}$.

\begin{figure*}[th!]
	\centering
 	\includegraphics[width=0.7\linewidth]{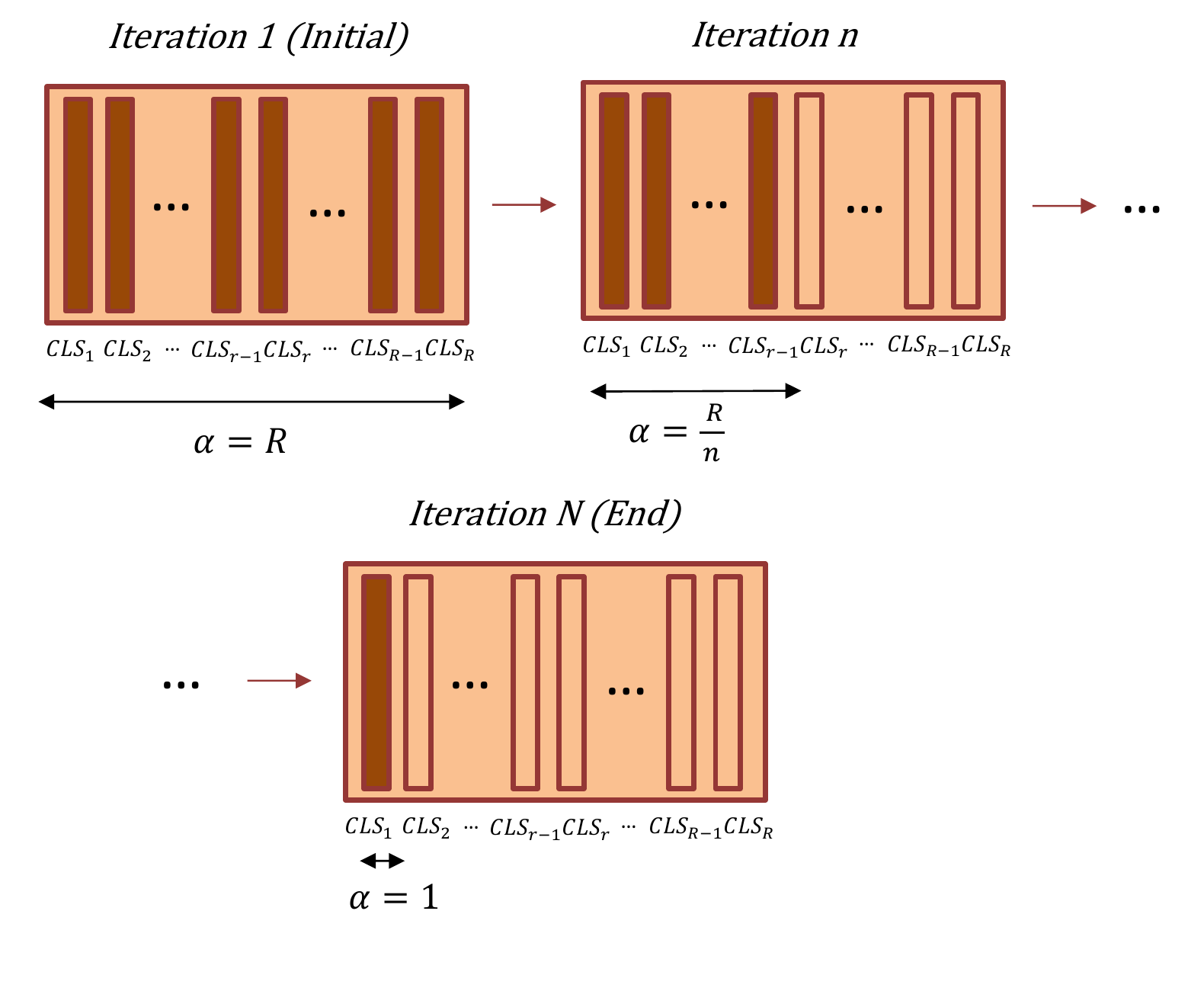}
    \caption{Evolution of the number of vectors in the token matrix that are available for sampling from the beginning of the training process (iteration 1) to the final iteration (iteration N). In each iteration, the dark vectors represent the enabled class tokens, while the light vectors are the disabled tokens.} 
    \label{fig1}
\end{figure*}

To introduce the class token in the system, one trainable vector parameter with the dimension of the feature vectors is defined when the network is initialized.
For each batch, it is replicated and concatenated at the end of each input feature sequence in the training batch as an additional token.
Hence, a single shared vector is trained to learn the final embedding representation.
%

In this work, we propose the use of a new sampling approach \cite{blundell2015weight}, and instead of having a single class token shared for the whole batch, we assume this sensitive parameter is the result of sampling a list of several vectors to be selected during the training by sampling them.
In order to do that, we define a matrix of $R$ vectors ($Token \ Matrix$) and sample it to take one of them for each example in the batch introducing uncertainty in the class token ($CLS  \ Token$). 
However, the use of this approach leads to a complex and slower evaluation process since a sampling inference would have to be carried out to obtain the representations.
For this reason, to avoid making the sampling inference, we have scheduled a forced reduction of the available vectors in the $Token \ Matrix$ throughout the training process.
Thus, at the end of this process, only one weight is different to zero, and the class token vector parameter is fixed. 
This strategy allows us to start the training ($Iteration \ 1$) with a matrix of several vectors to sample from and, gradually, we reduce the number of vectors as the training progresses to finish ($Iteration \ N$) with only one as the original class token as Fig.\ref{fig1} depicts.
Therefore, the training leads the system progressively to focus the relevant information on the first vector of the matrix.
In addition, using this sampling approach, the system is trained to capture the uncertainty introduced by initially having a $Token \ Matrix$ with $R$ vectors to combine with the training batch data.
Thus, each example from the batch is combined with a random vector from the matrix which is reduced in size after each epoch until only one vector remains at the end, so more variability has to be modelled which helps to improve the robustness of the system.
%

\begin{figure*}[th!]
	\centering
 	\includegraphics[width=0.8\linewidth]{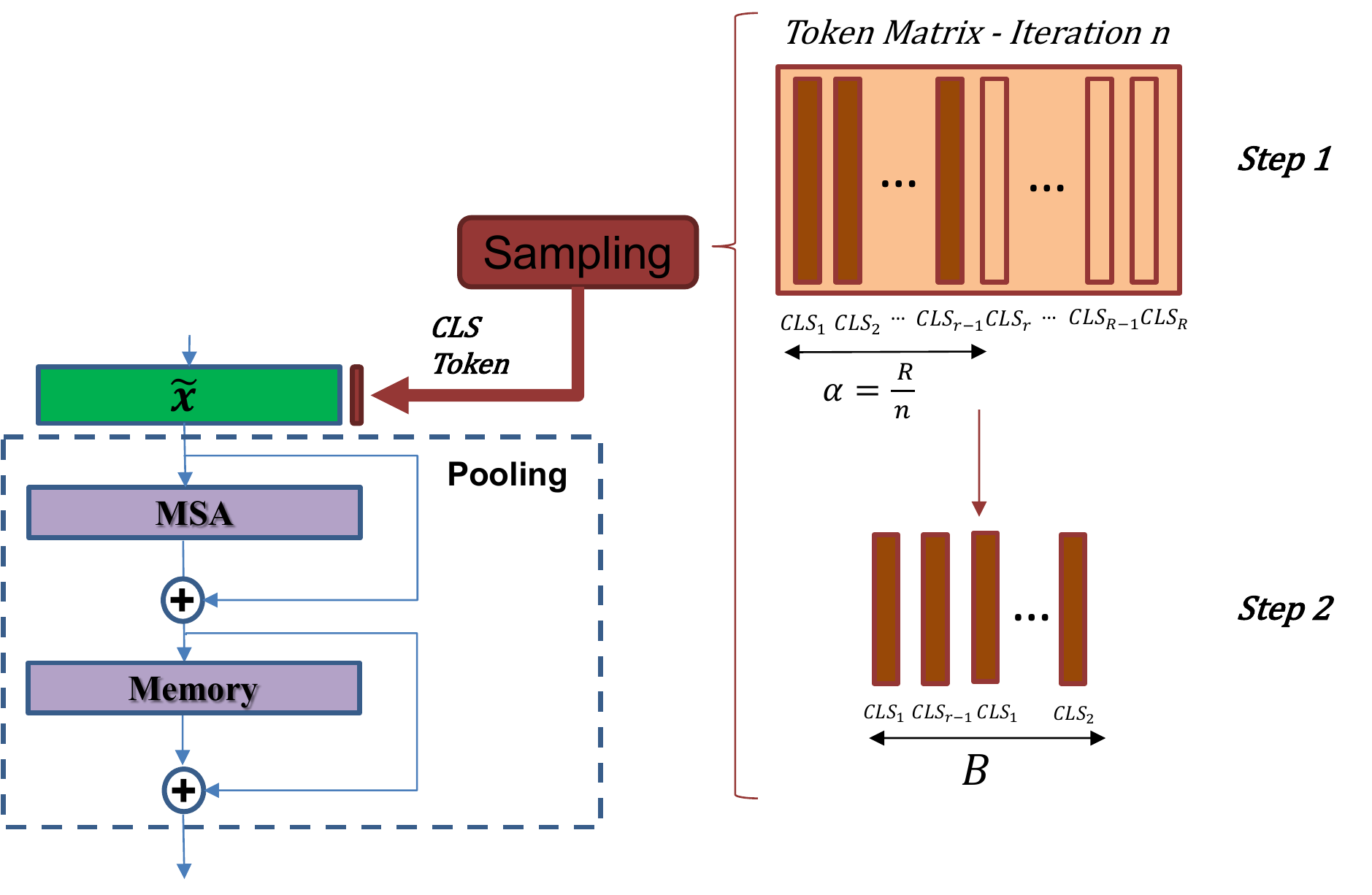}
    \caption{Example of the sampling steps in iteration n of the training process. In Step 1, the available vectors of the token matrix in that iteration are defined and the random indeces of batch size ($B$) are calculated. In Step 2, the class tokens are selected and added to the input of MSA layer.} 
    \label{fig2}
\end{figure*}

\begin{algorithm*}[t!]
\SetAlgoLined
\KwIn{Input examples $\mathcal{X}$, batch size $B$, examples of batch $x$, the number of layers $L$, the total number of tokens to sample $R$, and the number of epochs $N$ }
\vspace{0.1cm}

\textbf{1. Define the vector with the number of sample vectors $\alpha$ available to select each epoch:}\

\vspace{0.1cm}

$\mathcal{\alpha} = (\alpha_{1}, ..., \alpha_{n}, ..., \alpha_{N})$, $\alpha_{n}=R, ..., 1,$ with $R$ $\in$ $\mathbb{R}$ 

\vspace{0.1cm}

\textbf{2. Define the random matrix of class token vectors:}\

$Token Matrix = random\_matrix(R)$

\vspace{0.1cm}
\For{ $n=1$ \KwTo $N$ }{
\vspace{0.1cm}
\For{ $x \in \mathcal{X}$ }{
\vspace{0.1cm}
\textbf{3. Sampling Process}

\vspace{0.1cm}

\textbf{3.1 Step 1,  every update $B$ integer indexes are randomly generated from the available $\alpha_{n}$ vectors:}\

$inds$=$random\_integer$($\alpha_{n}$,$B$)

\vspace{0.1cm}

\textbf{3.2 Step 2, the correspondent tokens are selected:}\
    
$ CLS_{Token} =Token Matrix[inds]$
    
\vspace{0.1cm}
    \textbf{4. Network Training}
    
    \vspace{0.1cm}
    
    \textbf{4.1 Step 1, the class token is concatenated with the input to the MSA layer: }\
    
    \vspace{0.1cm}
    
    $x_{l}=[ x||CLS_{Token}]$
    
    \vspace{0.1cm}
    
    \textbf{4.2 Step 2, the new input is introduced to the first MSA layer in the pooling part and the L layers are applied:}\
    \vspace{0.1cm}

    \For{ $l=1$ \KwTo $L$ }{
        $x_{l}^{'}=x_{l}+MSA(x_{l})$
    
        $x_{l}=x_{l}^{'}+Memory(x_{l}^{'})$
    }
    
    \vspace{0.1cm}
    
    \textbf{4.3 Step 3, the state at the output of the last layer in the pooling block of the class token is used as final representation:}\
    
    $x_{CLS}= x_{l}^{end}$
}    
}

\caption{Algorithm for sampling class token and introducing it before the pooling part.}
\label{alg:1}
\end{algorithm*}

To carry out this process, we define the following vector, which indicates to the neural network the number of tokens available at each iteration of the training process: 

\begin{equation}
\mathcal{\alpha} = (\alpha_{1}, ..., \alpha_{n}, ..., \alpha_{N}), \alpha_{n}=R, ..., 1, with \ R\in \mathbb{R} 
\end{equation}
where $R$ is the number of tokens defined in the matrix, and $N$ is the total number of iterations for training process.
Among the number of tokens available at each iteration, a random selection of the batch size is made to select the index of the vectors.
These vectors are selected from the distribution ($Token \ Matrix$) and used as class tokens ($CLS  \ Token$) in the batch to concatenate to the input of the first MSA layer. 
The overall process is described in Algorithm \ref{alg:1}. 
Besides, Fig.\ref{fig2} shows a graphical example of how this sampling process is made in an intermediate iteration ($Iteration \ n$).
In this graphical explanation, it can be observed how in an intermediate iteration the number of vectors has been reduced forcing the network to put the relevant information to represent the utterances in the vectors still available.
%


\section{Knowledge Distillation with Tokens}
\begin{figure*}[th!]
    \centering
    \includegraphics[width=0.8\linewidth]{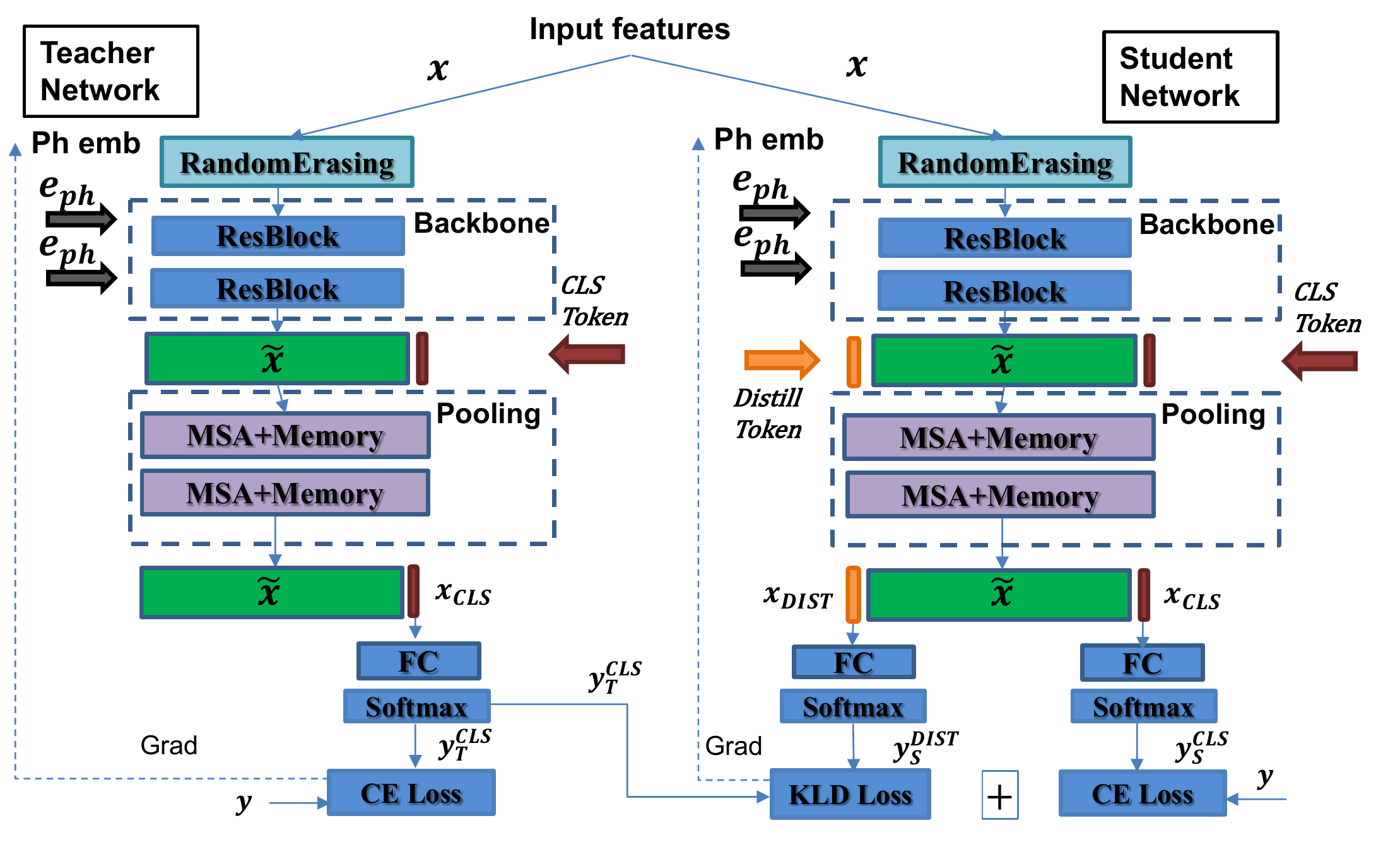}
    \caption{Teacher-student architecture used to create the system, where the dashed line indicates the process of backpropagation of the gradients. Both networks are employed to train while for testing, the student network is the only used.} 
    \label{fig3}
\end{figure*}

Motivated by the benefits obtained when the training databases are not very large with Teacher-Student architecture based on CNNs \cite{mingote2020bdk}, we have implemented this architecture using two transformer networks as Fig.\ref{fig3} depicts.
Using a Bayesian approach similar to \cite{Korattikara2015BayesianKnowledge}, the teacher-student architecture allows providing robustness to the system.
In this approach, the teacher and student networks are trained at the same time, unlike previous works \cite{shen2018feature,shen2019interactive} in which the teacher network is usually a pre-trained model to reduce complexity.
Whether the teacher network had been a frozen model, negative training examples that obtain high posterior values in the teacher network would be learned as positive examples by the student network.
Besides, different sources of distortion are applied to each of the input signals of both networks, so we have employed a data augmentation method called Random Erasing (RE) \cite{zhong2020random} to provide more variability to the input training data.
With this kind of architecture, the teacher network has to predict augmented unseen data and the student network tries to mimic the label predictions produced by the teacher network using the class token output.
This training strategy allows the student network to capture the variability in the predictions produced by the first network and model this uncertainty in the parameters during the training process.
However, inspired by \cite{touvron2020training}, we have also included an extra learnable token in the student network which is known as Distillation Token ($Distill \ Token$).
The introduction of this extra token allows to implement of a multi-objective optimization by using the class token to reproduce the true label while the distillation token is trained to mimic the predictions of the teacher network.
To achieve this, the Kullback-Leibler Divergence (KLD) loss between the student and teacher distributions is minimized.
The KLD loss can be formulated as,

\begin{equation}
\resizebox{0.825\linewidth}{!}
{$KLD=-{\sum_{i=1}^{I}\sum_{j=1}^{J}{p_{T}(y_{i}^{cls}|x_{j})\cdot \log{(p_{S(y_{i}^{dist}|x_{j}))}}}+const}$},
\label{eq:loss}
\end{equation}
where $i$ and $j$ are the speaker and utterance indices, $x_{j}$ is the input signal, $p_{T}(y_{i}^{cls}|x_{j})$ is the output posterior probability of the label $y_{i}^{cls}$ from the class token of the teacher model, $p_{S}(y_{i}^{dist}|x_{j})$ is the output posterior probability of the label $y_{i}^{dist}$ from the distillation token of the student network for the same example, and $const$ is defined in \cite{Korattikara2015BayesianKnowledge}.
Hence, to train the teacher-student architecture showed in Fig.\ref{fig3}, we employ the following two loss expressions for teacher and student networks: 

\begin{equation}
Loss_{T}=CE(y_{T}^{cls},y),
\label{eq:losscompleta1}
\end{equation}

\begin{equation}
Loss_{S}=KLD(y_{S}^{dist},y_{T}^{cls})+CE(y_{S}^{cls},y),
\label{eq:losscompleta2}
\end{equation}
where $CE$ is the cross-entropy loss, $y_{S}^{cls}$ is the class token output from the student network, and $y$ are the ground truth labels.

\section{System Description}
In this section, we describe the system architecture used in this work for text-dependent SV.
Fig.\ref{fig3} depicts this architecture where a teacher-student approach is employed.
Both architectures follow the structure described in \cite{Mingote2021icassp} with the same backbone and pooling parts.
The backbone is based on two Residual Network (RN) \cite{he2016deep} blocks with three layers each block.
Additionally, these architectures need embeddings with positional information to help guiding the attention mask in the MSA layers.
In this work, these embeddings ($e_{ph}$) are extracted by a phonetic classifier network instead of using temporal position information \cite{vinals2019phonetically}. 
For the pooling part, two MSA layers of 16 heads combined with two memory layers are employed.
Moreover, before the first MSA layer, the class token is concatenated to the input.
In the case of the student network, the distillation token is also included.
Thanks to the self-attention mechanism, these tokens learn to obtain a global representation for each utterance without applying the global average pooling. 
These representations, similar to supervector, are more convenient for text-dependent SV task since these global representations do not neglect the sequence order and are obtained automatically thanks to the self-attention mechanism. 
So external alignment mechanisms are not necessary to obtain them as in \cite{Mingote2018,mingote2019supervector,mingote2020optimization}, where GMM or HMM posterior probabilities are needed to align speech frames to supervectors. 
%
Besides, the use of memory layers increases the amount of knowledge obtained by the network that can be stored.  
After training the system, a cosine similarity over the token representations is applied to perform the verification process.
Note that this kind of system based on teacher-student consists of training of two architectures at the same time.
Therefore, this process may involve a higher computational cost.
However, during inference, only the student network is employed to extract the embeddings, so there is no extra inference time.

\section{Experimental Setup}
\subsection{Datasets}
For the experiments, two text-dependent speaker verification datasets have been employed.
The first set of experiments has been reported on the RSR2015 text-dependent speaker verification dataset \cite{Larcher2014Text-dependentRSR2015}.
This dataset comprises recordings from 157 males and 143 females. 
For each speaker, there are 9 sessions with 30 different phrases.
This data is divided into three speaker subsets: background (bkg), development (dev) and evaluation (eval). 
In this paper, we develop our experiments with Part II, which is composed of short control command with a strong overlap of lexical content, and we employ only the bkg data for training. 
The eval part is used for enrollment and trial evaluation. 
This dataset has three evaluation conditions, but in this work, the most challenging, which is the Impostor-Correct case, is the only one that has been evaluated and employed in the text-dependent SV.
Note that there are other systems that obtain relevant results for this dataset, similar to those presented below.
Nevertheless, such systems are based on traditional models such as Hidden Markov Models (HMMs) \cite{Larcher2014Text-dependentRSR2015,das2018compensating} or neural network architectures focused on two different streams for speaker and utterance information \cite{liu2019unified,liu2020speaker}. 

The second dataset used is the DeepMine database \cite{deepmine2019asru}.
This corpus consists of three different parts of which we employ the files selected for the Short-duration Speaker Verification (SdSV) Challenge 2020 \cite{sdsvc2020plan} from Part 1.
Part 1 is the text-dependent part which is composed of 5 Persian and 5 English phrases and contains 963 females and males speakers.
This data is divided into two subsets: train with 101.063 audio files and evaluation with 69.542 audio files.
Finally, the phonetic classification network  \cite{vinals2019phonetically} has been trained using LibriSpeech \cite{panayotov2015librispeech} to extract phonetic embeddings.
Unlike other works presented in the challenge \cite{lozano2020but,chen2020improving}, we have not used VoxCeleb 1 and 2 datasets \cite{nagrani2017voxceleb,chung2018voxceleb2} in the neural network training process.
Motivated by the fact that in some situations and applications is required the implementation of custom systems with the few available in domain-data.
For this reason, we have developed systems only with the in-domain data.

\subsection{Experimental Description}
To carry out the experiments with the RSR2015 dataset, a set of features composed of 20 dimensional Mel-Frequency Cepstral Coefficients (MFCC) with their derivates are employed as input.
While for the experiments using the DeepMine dataset, we have employed a feature vector based on mel-scale filter banks.
With this feature extractor, we obtain two log filter banks of sizes 24 and 32, which are concatenated with the log energy to obtain a final input dimension of 57.
Moreover, phonetic embeddings of 256 dimensions have been used as positional information.
As the optimizer for the experiments in this work, the Adam optimizer is employed with a learning rate that increases from $10^{-3}$ to $5*10^{-3}$ during 60 epochs and then decays from $5*10^{-3}$ to $10^{-4}$.
In addition, training data is fed into the systems with a minibatch size of 32. 

\section{Results}
In this paper, two sets of experiments have been carried out to evaluate the proposals with both databases.
We compare the different approaches to obtain the representations with a single neural network using the same architecture as the teacher network: the use of the traditional global average pooling ($AVG$), the attentive pooling ($ATT$) and the introduction of the learnable class token ($CLS$).
%
%
For the class token approach, we evaluate our proposal of sampling a matrix of $R$ vectors and reducing it until having a single vector ($Sampling$).
This parameter is also swept for different values of $R$, including $R=1$ that corresponds to the original idea of having a single token and repeating it.
Moreover, we analyze the effect produced by the fact of using a teacher-student architecture with an extra distillation token ($CLS-DIST$).

In order to evaluate these experiments, we have measured the performance using three metrics. 
Equal Error Rate (EER) which measures the discrimination ability of the system.
NIST 2008 and 2010 minimum Detection Cost Functions (DCF08, DCF10) \cite{nist2008,nist2010} which measure the cost of detection errors in terms of a weighted sum of false alarm and miss probabilities for a decision threshold, and a priori probability.

\subsection{Class Token Study}
A first set of experiments was performed to compare the use of a class token to obtain global utterance descriptors with the use of a global average pooling method or the attentive pooling proposed in \cite{desplanques2020ecapa}.
Thus, we study the two approaches to introduce this vector explained during this work and the effect of the number of vectors chosen for the sampling approach.

Table \ref{tab:table1} presents EER, DCF08 and DCF10 results for the experiments with RSR2015-part II dataset.
Regardless of the number of vectors in the sampling for class tokens, if we apply our proposed strategy to introduce the tokens with a sampling alternative, the obtained performance is better.
In addition, the results show how employing a learnable token outperforms the use of an average embedding or an attentive pooling embedding.
Note that the token is trained through self-attention and keeping the temporal structure to obtain a global utterance representation, while the average embedding neglects this information that is relevant to the SV task. 
As we can also observe with the sweep of $R$ value, the use of several vectors to create the token matrix is better than using a single vector and repeating it for the whole batch, which corresponds to the original way of applying this approach.
The case of having a single vector and repeat it corresponds with the experiments with $R=1$.
However, when the number of available tokens is too large, the performance begins to degrade.
This degradation could be caused by the introduction of too much variability that the system is not able to model as the architectures employed are not so large, which means that there are a limited number of different tokens to carry out the training process.

\begin{table*}[th!]
  	\caption{Experimental results on RSR2015 Part II \cite{Larcher2014Text-dependentRSR2015} eval subset, showing EER\%, DCF08 and DCF10. These results were obtained to compare the different approaches to obtain the representations: average, attentive or sampling strategies. }
  	\label{tab:table1}
  	\centering
  	\resizebox{1.0\textwidth}{!} {
  	\begin{tabular}{c c c c c c |c c c |c c c}
    \hline    
    \multicolumn{3}{c}{\textbf{Architecture}}&
    \multicolumn{3}{c}{\textbf{Female}}&
    \multicolumn{3}{c}{\textbf{Male}}&
    \multicolumn{3}{c}{\textbf{Female + Male }}\\
    \cline{1-3}
    \multicolumn{1}{c}{\textbf{Type}}&
    \multicolumn{1}{c}{\textbf{T/S}}&
    \multicolumn{1}{c}{\textbf{Sampling}}&
    \multicolumn{1}{c}{\textbf{EER\%}}&
    \multicolumn{1}{c}{\textbf{DCF08}}&
    \multicolumn{1}{c}{\textbf{DCF10}}&
    \multicolumn{1}{c}{\textbf{EER\%}}&
    \multicolumn{1}{c}{\textbf{DCF08}}&
    \multicolumn{1}{c}{\textbf{DCF10}}&
    \multicolumn{1}{c}{\textbf{EER\%}}&
    \multicolumn{1}{c}{\textbf{DCF08}}&
    \multicolumn{1}{c}{\textbf{DCF10}}\\
    \hline
    AVG&no&$-$&$4.64$&$0.228$&$0.669$&$4.92$&$0.244$&$0.716$&$4.79$&$0.237$&$0.706$\\
    \hline
    ATT&no&$-$&$4.07$&$0.188$&$0.618$&$4.44$&$0.211$&$0.615$&$4.53$&$0.213$&$0.646$\\
    \hline
    CLS&no&R=1&$3.71$&$0.174$&$0.580$&$4.27$&$0.215$&$0.679$&$4.12$&$0.201$&$0.634$\\
    $ $&$ $&R=50&$3.37$&$0.169$&$0.580$&$4.04$&$0.199$&$0.601$&$3.75$&$0.187$&$0.606$\\
    $ $&$ $&R=100&$\textbf{3.33}$&$\textbf{0.158}$&$\textbf{0.552}$&$\textbf{3.68}$&$\textbf{0.182}$&$\textbf{0.552}$&$\textbf{3.57}$&$\textbf{0.173}$&$\textbf{0.565}$\\
    $ $&$ $&R=200&$3.55$&$0.171$&$0.562$&$4.09$&$ 0.199$&$0.607$&$3.86$&$0.189$&$0.587$\\
    \hline
	\end{tabular}}
\end{table*}

In Table \ref{tab:table2}, the results obtained in DeepMine-part 1 database are shown.
Unlike the other dataset, the training data in DeepMine is larger, which indicates that the lack of data is not so critical to train a powerful and robust system.
Therefore, the replacement of the average embedding or attentive pooling embedding by a class token improves the performance only slightly.
Besides, the sweep of $R$ value shows that the evolution of the female and male results separately do not follow the same trend as occurs in the RSR-Part II results.

\begin{table*}[th!]
  	\caption{Experimental results on DeepMine \cite{deepmine2019asru} eval subset, showing EER\%, DCF08 and DCF10. These results were obtained to compare the different approaches to obtain the representations: average, attentive or sampling strategies. }
  	\label{tab:table2}
  	\centering
  	\resizebox{1.0\textwidth}{!} {
  	\begin{tabular}{c c c c c c |c c c |c c c}
    \hline    
    \multicolumn{3}{c}{\textbf{Architecture}}&
    \multicolumn{3}{c}{\textbf{Female}}&
    \multicolumn{3}{c}{\textbf{Male}}&
    \multicolumn{3}{c}{\textbf{Female + Male }}\\
    \cline{1-3}
    \multicolumn{1}{c}{\textbf{Type}}&
    \multicolumn{1}{c}{\textbf{T/S}}&
    \multicolumn{1}{c}{\textbf{Sampling}}&
    \multicolumn{1}{c}{\textbf{EER\%}}&
    \multicolumn{1}{c}{\textbf{DCF08}}&
    \multicolumn{1}{c}{\textbf{DCF10}}&
    \multicolumn{1}{c}{\textbf{EER\%}}&
    \multicolumn{1}{c}{\textbf{DCF08}}&
    \multicolumn{1}{c}{\textbf{DCF10}}&
    \multicolumn{1}{c}{\textbf{EER\%}}&
    \multicolumn{1}{c}{\textbf{DCF08}}&
    \multicolumn{1}{c}{\textbf{DCF10}}\\
    \hline
    AVG&no&$-$& $3.92$& $0.135$& $0.411$ & $3.02$& $0.137$& $0.676$ &$3.58$&$0.136$&$0.521$ \\
    \hline
    ATT&no&$-$&$5.73$&$0.193$&$0.468$&$5.39$&$0.210$&$0.664$&$5.60$&$0.200$&$0.551$\\
    \hline
    CLS&no&R=1& $3.81$& $0.128$& $0.389$ & $3.32$& $0.143$& $0.697$ &$3.60$&$0.134$&$0.520$\\
    $ $&$ $&R=50& $3.92$& $0.131$& $0.393$ & $3.19$& $0.140$& $0.668$ &$3.62$&$0.134$&$0.519$\\
    $ $&$ $&R=100& $\textbf{3.69}$& $\textbf{0.124}$& $\textbf{0.379}$ & $3.09$& $0.137$& $0.658$ &$\textbf{3.43}$&$\textbf{0.129}$&$\textbf{0.505}$\\
    $ $&$ $&R=200& $3.89$& $0.133$& $0.417$ & $\textbf{2.92}$& $\textbf{0.133}$& $\textbf{0.655}$ &$3.50$&$0.133$&$0.521$\\
    \hline
	\end{tabular}}
\end{table*}

\subsection{Effect of Knowledge Distillation using Tokens}
In this section, we analyze the effect of introducing an approach based on Knowledge Distillation philosophy which consists of a teacher-student architecture.
Furthermore, in this approach, an extra distillation token ($CLS-DIST$) is incorporated \cite{touvron2020training}. 
This approach has been employed to compare the performance obtained in the case of the average global pooling as well as in the proposed sampling approach to use the class token.
In this second case, we have developed the teacher-student architecture using the $R$ value of the best configuration obtained in the previous section, and also, the case of $R=1$ as it is the usual way to apply this class token approach in the literature.  

Results of these experiments in RSR-Part II are shown in Table \ref{tab:table3}.
Regardless of the kind of approach to obtain the representations used, we can observe that the use of an architecture based on a teacher-student approach improves the robustness and achieves better performance in all the alternatives to extract the representations.
Moreover,  the best performance is obtained applying our proposed strategy to introduce the tokens with a sampling alternative with more than a single vector.

\begin{table*}[th!]
  	\caption{Experimental results on RSR2015 Part II \cite{Larcher2014Text-dependentRSR2015} eval subset, showing EER\%, DCF08 and DCF10. These results were obtained to compare the use of a teacher-student architecture for the different approaches to obtain the representations: average or sampling strategies.}
  	\label{tab:table3}
  	\centering
  	\resizebox{1.0\textwidth}{!} {
  	\begin{tabular}{c c c c c c |c c c |c c c}
    \hline    
    \multicolumn{3}{c}{\textbf{Architecture}}&
    \multicolumn{3}{c}{\textbf{Female}}&
    \multicolumn{3}{c}{\textbf{Male}}&
    \multicolumn{3}{c}{\textbf{Female + Male }}\\
    \cline{1-3}
    \multicolumn{1}{c}{\textbf{Type}}&
    \multicolumn{1}{c}{\textbf{T/S}}&
    \multicolumn{1}{c}{\textbf{Sampling}}&
    \multicolumn{1}{c}{\textbf{EER\%}}&
    \multicolumn{1}{c}{\textbf{DCF08}}&
    \multicolumn{1}{c}{\textbf{DCF10}}&
    \multicolumn{1}{c}{\textbf{EER\%}}&
    \multicolumn{1}{c}{\textbf{DCF08}}&
    \multicolumn{1}{c}{\textbf{DCF10}}&
    \multicolumn{1}{c}{\textbf{EER\%}}&
    \multicolumn{1}{c}{\textbf{DCF08}}&
    \multicolumn{1}{c}{\textbf{DCF10}}\\
    \hline
    AVG&no&$-$&$4.64$&$0.228$&$0.669$&$4.92$&$0.244$&$0.716$&$4.79$&$0.237$&$0.706$\\
    $ $&yes&$-$&$3.52$&$0.170$&$0.587$&$3.78$&$0.186$&$0.579$&$3.74$&$0.185$&$0.602$\\
    \hline
    CLS&no&R=1&$3.71$&$0.174$&$0.580$&$4.27$&$0.215$&$0.679$&$4.12$&$0.201$&$0.634$\\
    CLS-DIST&yes&R=1&$3.01$&$0.148$&$0.548$&$3.40$&$0.173$&$0.557$&$3.31$&$0.167$&$0.558$\\
    \hline
    CLS&no&R=100&$3.33$&$0.158$&$0.552$&$3.68$&$0.182$&$0.552$&$3.57$&$0.173$&$0.565$\\
    CLS-DIST&yes&R=100&$\textbf{2.47}$&$\textbf{0.122}$&$\textbf{0.414}$&$\textbf{2.83}$&$\textbf{0.138}$&$ \textbf{0.463}$&$\textbf{2.68}$&$\textbf{0.133}$&$\textbf{0.443}$\\
    \hline
	\end{tabular}}
\end{table*}

On the other hand, Table \ref{tab:table4} presents the performance of systems with DeepMine-part 1.
In this case, the results show that the application of only the teacher-student architecture does not improve the systems.
However, the use of the teacher-student architecture and the extra distillation token ($CLS-DIST$), combined with the sampling strategy with several token vectors also allows achieving a more robust system and a significant improvement in the results. 

\begin{table*}[th!]
  	\caption{Experimental results on DeepMine \cite{deepmine2019asru} eval subset, showing EER\%, DCF08 and DCF10. These results were obtained to compare the use of a teacher-student architecture for the different approaches to obtain the representations: average or sampling strategies. }
  	\label{tab:table4}
  	\centering
  	\resizebox{1.0\textwidth}{!} {
  	\begin{tabular}{c c c c c c |c c c |c c c}
    \hline    
    \multicolumn{3}{c}{\textbf{Architecture}}&
    \multicolumn{3}{c}{\textbf{Female}}&
    \multicolumn{3}{c}{\textbf{Male}}&
    \multicolumn{3}{c}{\textbf{Female + Male }}\\
    \cline{1-3}
    \multicolumn{1}{c}{\textbf{Type}}&
    \multicolumn{1}{c}{\textbf{T/S}}&
    \multicolumn{1}{c}{\textbf{Sampling}}&
    \multicolumn{1}{c}{\textbf{EER\%}}&
    \multicolumn{1}{c}{\textbf{DCF08}}&
    \multicolumn{1}{c}{\textbf{DCF10}}&
    \multicolumn{1}{c}{\textbf{EER\%}}&
    \multicolumn{1}{c}{\textbf{DCF08}}&
    \multicolumn{1}{c}{\textbf{DCF10}}&
    \multicolumn{1}{c}{\textbf{EER\%}}&
    \multicolumn{1}{c}{\textbf{DCF08}}&
    \multicolumn{1}{c}{\textbf{DCF10}}\\
    \hline
    AVG&no&$-$& $3.92$& $0.135$& $0.411$ & $3.02$& $0.137$& $0.676$ &$3.58$&$0.136$&$0.521$ \\
    $ $&yes&$-$&$4.07$& $0.135$& $0.401$ & $3.04$& $0.141$& $0.646$ &$3.65$&$0.138$&$0.501$\\
    \hline
    CLS&no&R=1& $3.81$& $0.128$& $0.389$ & $3.32$& $0.143$& $0.697$ &$3.60$&$0.134$&$0.520$\\
    CLS-DIST&yes&R=1&$3.80$& $0.131$& $0.395$ & $3.25$& $0.144$& $\textbf{0.621}$ &$3.57$&$ 0.135$&$0.494$\\
    \hline
    CLS&no&R=100& $3.69$& $0.124$& $\textbf{0.379}$ & $3.09$& $0.137$& $0.658$ &$3.43$&$0.129$&$0.505$\\
    CLS-DIST&$yes$&R=100&$\textbf{3.51}$& $\textbf{0.122}$& $0.385$ & $\textbf{2.68}$& $\textbf{0.122}$& $0.652$ &$\textbf{3.19}$&$\textbf{0.123}$&$\textbf{0.492}$\\
    \hline
	\end{tabular}}
\end{table*}

\subsection{Analysis of Class Token Self-Attention Representations}
\begin{figure*}[th!]
    \centering
	\subfigure[Call sister]{
 	\includegraphics[width=0.485\linewidth]{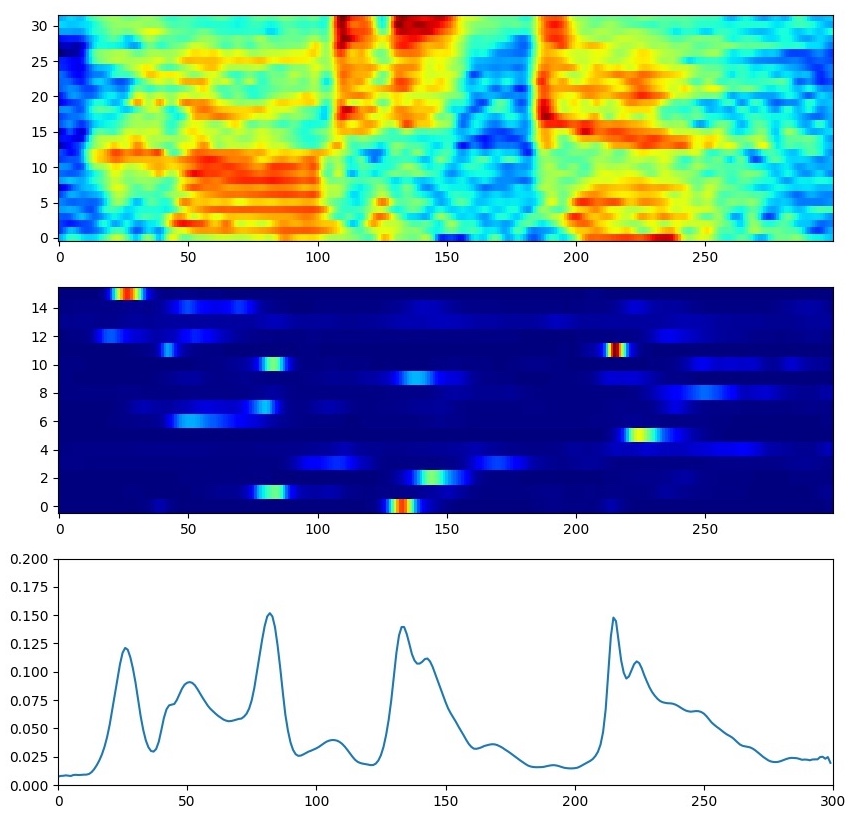}
	\label{fig4:a}}
	\centering
	\subfigure[Call brother]{
 	\includegraphics[width=0.481\linewidth]{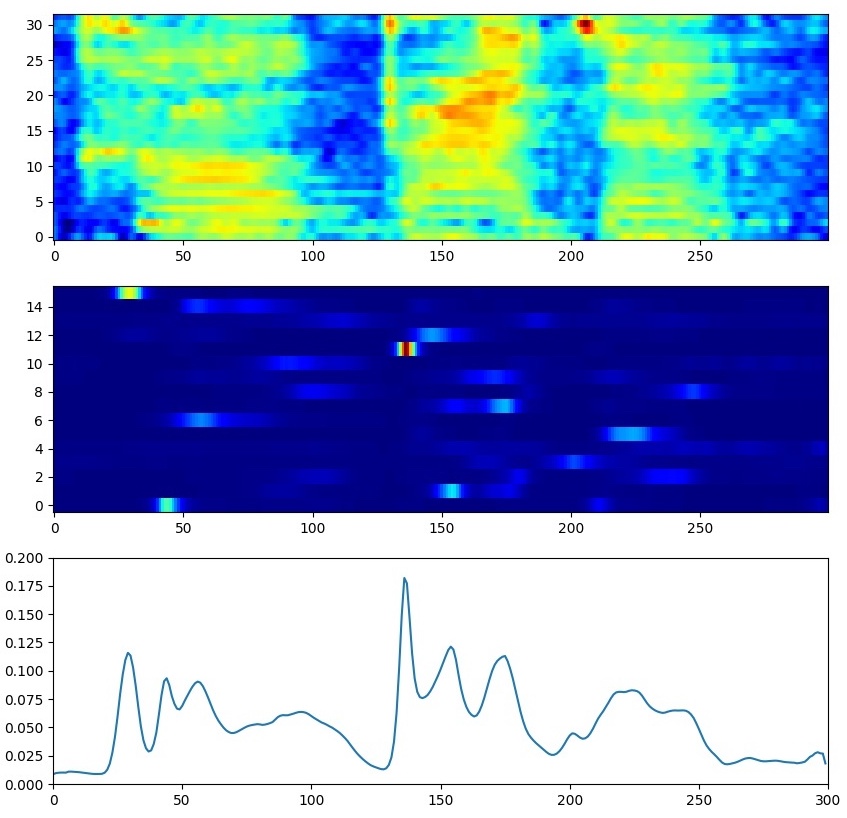}
	\label{fig4:b}}
    \caption{Visualizing two examples of different phrases of RSR-Part II which are pronounced by the same speaker. In both cases, three representations are presented. The figure on top shows the spectrogram of each phrase. In the middle, the attention weights learnt by the class token for each of the 16 heads in the last MSA layer are depicted. Finally, the plot on bottom is the sum of the rows of the previous weight attention matrix. } 
    \label{fig4}
\end{figure*}

\begin{figure*}[th!]
    \centering
	\subfigure[OK Google]{
 	\includegraphics[width=0.481\linewidth]{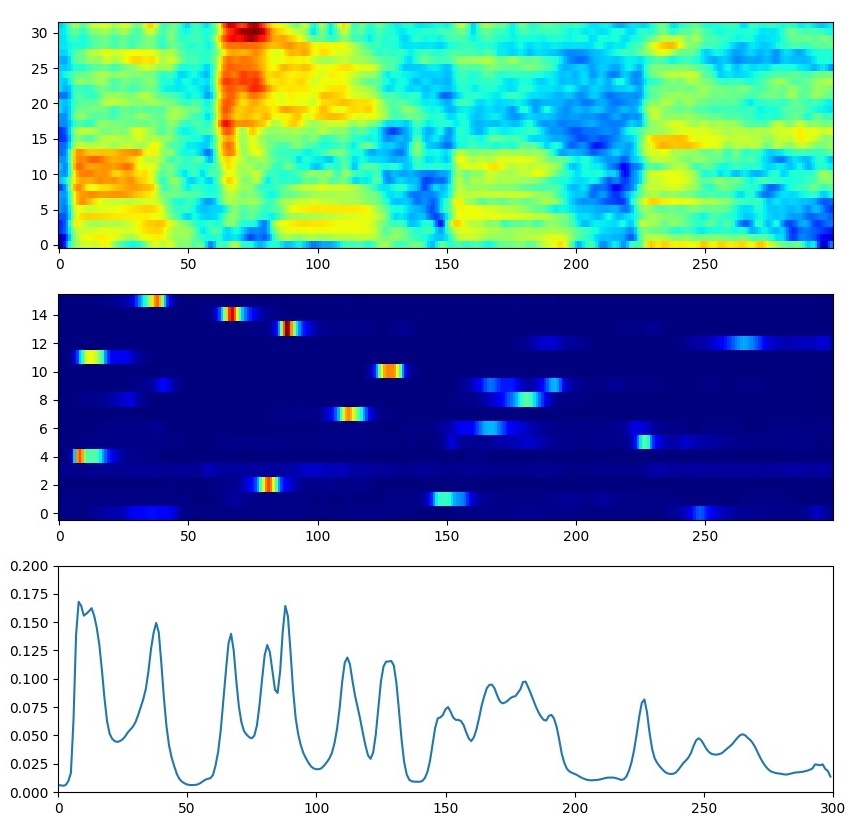}
	\label{fig5:a}}
	\centering
	\subfigure[OK Google]{
 	\includegraphics[width=0.484\linewidth]{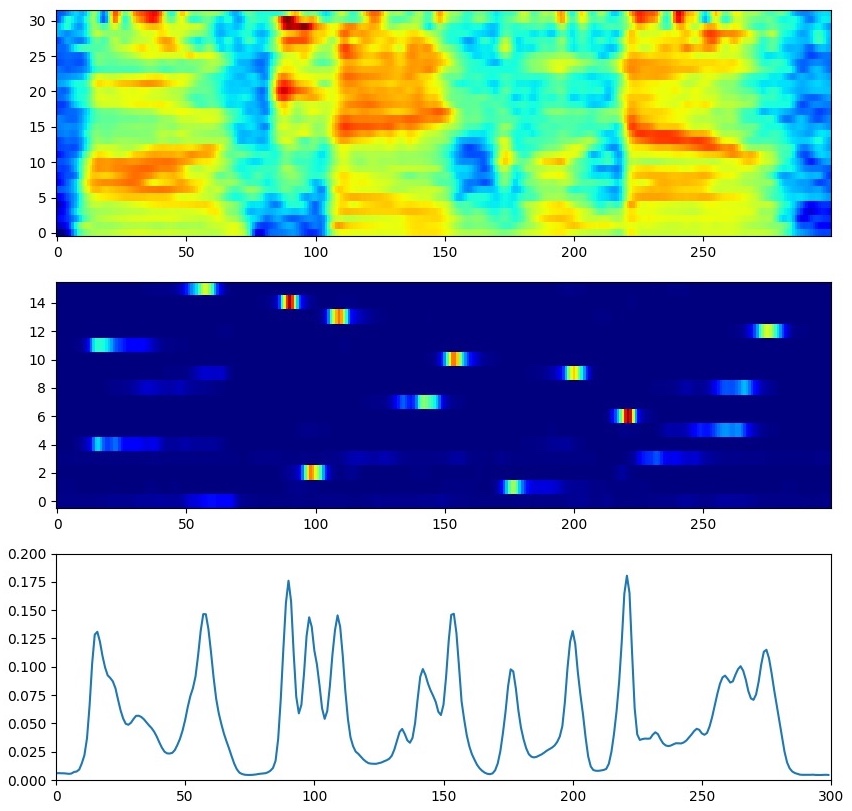}
	\label{fig5:b}}
    \caption{Visualizing two examples of the same phrase of DeepMine which are pronounced by different speakers. In both cases, three representations are presented. The figure on top shows the spectrogram of each phrase. In the middle, the attention weights learnt by the class token for each of the 16 heads in the last MSA layer are depicted. Finally, the plot on bottom is the sum of the rows of the previous weight attention matrix.} 
    \label{fig5}
    \vspace{-0.3cm}
\end{figure*}

In view of the relevant results obtained, we have also conducted an analysis to interpret what the self-attention matrix $A$ is learning in each system.
To perform this analysis, we have employed the system with the best performance from each database, and within these systems, the last MSA layer of the student model has been selected to make the representations.
In addition, we have chosen different utterances to analyze in Fig.\ref{fig4} and Fig.\ref{fig5}.
For each utterance, three figures are plotted: the spectrogram of the utterance, the matrix of attention weights corresponding to the class token for each of the 16 heads of the MSA layer, and the sum of the weights of these class token attentions.

In Fig.\ref{fig4}, two examples of utterances of different phrases (\textit{``Call sister''}, \textit{``Call brother''}) pronounced by the same speaker are shown. 
These examples are obtained from the evaluation set of the RSR-Part II database.
Whether we look in the middle and bottom figures, we can observe the relevant information learned by the self-attention weights to correctly determine the phrase and speaker of each utterance using the class token.
Note that these two phrases of example begin exactly the same with the word $Call$, so focusing on the beginning of the figures, we observe how the self-attention gives similar relevance in both cases to the areas of same phonemes.
Moreover, we can also see that the weights do not pay attention to the area at the beginning and end of the utterances that correspond to moments of silence. 

Fig.\ref{fig5} represents two examples of utterances of the same phrase (\textit{``OK Google''}) pronounced by different speakers. 
In this case, the examples are obtained from the evaluation set of the DeepMine database.
Note that since these figures are of the same phrase, self-attention is focused on the same areas, but different relevance is given to some of them.
Besides, the effect of not focusing on the beginning and end of the utterance also occurs in these examples.

\section{Conclusion}
\vspace{-0.2cm}
In this paper, we have presented a novel approach for the SV task. 
This approach is based on the use of a learnable class token to obtain a global utterance descriptor instead of employing the average pooling.
Moreover, we have developed an alternative to create the class token with a sampling strategy that introduces uncertainty that helps to generalize better.
Apart from the previous approach, we have also employed a teacher-student architecture combined with an extra distillation token to develop a more robust system.
Using this architecture, the distillation token in the student network learns to replicate the predictions from the teacher network.
Both proposals were evaluated in two text-dependent SV databases.
Results achieved show in RSR2015-part II that each of the approaches introduced to obtain a robust system and reduce potential underperformance due to the lack of data improves the overall performance.
However, in DeepMine-part 1, the results obtained  replacing only the average embedding by the class token present a small improvement, while the use of a teacher-student architecture achieves a great improvement and confirms the power of this kind of approach to train the systems.
\vspace{-0.2cm}

\section*{Credit Authorship Contribution Statement}
\vspace{-0.2cm}
\textbf{Victoria Mingote}: Conceptualization, Investigation, Methodology, Software, Writing- Original draft preparation; \textbf{Antonio Miguel}: Conceptualization, Investigation, Methodology, Software, Supervision, Writing- Reviewing and Editing.; \textbf{Alfonso Ortega}: Supervision, Writing- Reviewing and Editing; \textbf{Eduardo Lleida}: Supervision, Writing- Reviewing and Editing.

\section*{Acknowledgment}
This project has received funding from the European Union’s Horizon 2020 research and innovation programme under Marie Skłodowska-Curie Grant 101007666; in part by MCIN/AEI/10.13039/501100011033 and by the European Union “NextGenerationEU” / PRTR under Grant PDC2021-120846-C41, by the Spanish Ministry of Economy and Competitiveness and the European Social Fund through the grant PRE2018-083312, by the Government of Aragón (Grant Group T36\_20R), and by Nuance Communications, Inc.


\bibliography{mybibfile}

\begin{thebibliography}{10}
\expandafter\ifx\csname url\endcsname\relax
  \def\url#1{\texttt{#1}}\fi
\expandafter\ifx\csname urlprefix\endcsname\relax\def\urlprefix{URL }\fi
\expandafter\ifx\csname href\endcsname\relax
  \def\href#1#2{#2} \def\path#1{#1}\fi

\bibitem{Taigman2014DeepFace:Verification}
Y.~Taigman, M.~Yang, M.~Ranzato, L.~Wolf, {DeepFace: Closing the Gap to
  Human-Level Performance in Face Verification}, 2014 IEEE Conference on
  Computer Vision and Pattern Recognition (2014) 1701--1708\href
  {http://dx.doi.org/10.1109/CVPR.2014.220} {\path{doi:10.1109/CVPR.2014.220}}.

\bibitem{snyder2018x}
D.~Snyder, D.~Garcia-Romero, G.~Sell, D.~Povey, S.~Khudanpur, X-vectors: Robust
  dnn embeddings for speaker recognition, in: 2018 ICASSP, pp. 5329--5333.

\bibitem{Hoffer2015DeepNetwork}
E.~Hoffer, N.~Ailon, {Deep metric learning using triplet network}, Lecture
  Notes in Computer Science (including subseries Lecture Notes in Artificial
  Intelligence and Lecture Notes in Bioinformatics) 9370~(2010) (2015) 84--92.
\newblock \href {http://dx.doi.org/10.1007/978-3-319-24261-3{\_}7}
  {\path{doi:10.1007/978-3-319-24261-3{\_}7}}.

\bibitem{schroff2015facenet}
F.~Schroff, D.~Kalenichenko, J.~Philbin, {Facenet: A unified embedding for face
  recognition and clustering}, in: Proceedings of the IEEE conference on
  computer vision and pattern recognition, 2015, pp. 815--823.

\bibitem{snyder2016deep}
D.~Snyder, P.~Ghahremani, D.~Povey, D.~Garcia-Romero, Y.~Carmiel, S.~Khudanpur,
  Deep neural network-based speaker embeddings for end-to-end speaker
  verification, in: Spoken Language Technology Workshop (SLT), 2016 IEEE, IEEE,
  2016, pp. 165--170.

\bibitem{vaswani2017attention}
A.~Vaswani, N.~Shazeer, N.~Parmar, J.~Uszkoreit, L.~Jones, A.~N. Gomez,
  {\L}.~Kaiser, I.~Polosukhin, Attention is all you need, in: Advances in
  neural information processing systems, 2017, pp. 5998--6008.

\bibitem{kenton2019bert}
J.~D. M.-W.~C. Kenton, L.~K. Toutanova, Bert: Pre-training of deep
  bidirectional transformers for language understanding, in: Proceedings of
  NAACL-HLT, 2019, pp. 4171--4186.

\bibitem{dosovitskiy2020image}
A.~Dosovitskiy, L.~Beyer, A.~Kolesnikov, D.~Weissenborn, X.~Zhai,
  T.~Unterthiner, M.~Dehghani, M.~Minderer, G.~Heigold, S.~Gelly, et~al., {An
  image is worth 16x16 words: Transformers for image recognition at scale},
  arXiv preprint arXiv:2010.11929.

\bibitem{touvron2020training}
H.~Touvron, M.~Cord, M.~Douze, F.~Massa, A.~Sablayrolles, H.~J{\'e}gou,
  {Training data-efficient image transformers \& distillation through
  attention}, arXiv preprint arXiv:2012.12877.

\bibitem{locatello2020object}
F.~Locatello, D.~Weissenborn, T.~Unterthiner, A.~Mahendran, G.~Heigold,
  J.~Uszkoreit, A.~Dosovitskiy, T.~Kipf, {Object-Centric Learning with Slot
  Attention}, in: NeurIPS 2020, 2020.

\bibitem{india2019self}
M.~India, P.~Safari, J.~Hernando, Self multi-head attention for speaker
  recognition, Proc. Interspeech 2019 (2019) 4305--4309.

\bibitem{shim2022graph}
H.-j. Shim, J.~Heo, J.-h. Park, G.-h. Lee, H.-J. Yu, {Graph attentive feature
  aggregation for text-independent speaker verification}, in: ICASSP 2022-2022
  IEEE International Conference on Acoustics, Speech and Signal Processing
  (ICASSP), IEEE, 2022, pp. 7972--7976.

\bibitem{wang2022multi}
R.~Wang, J.~Ao, L.~Zhou, S.~Liu, Z.~Wei, T.~Ko, Q.~Li, Y.~Zhang, {Multi-View
  Self-Attention Based Transformer for Speaker Recognition}, in: ICASSP
  2022-2022 IEEE International Conference on Acoustics, Speech and Signal
  Processing (ICASSP), IEEE, 2022, pp. 6732--6736.

\bibitem{han2022local}
B.~Han, Z.~Chen, Y.~Qian, {Local Information Modeling with Self-Attention for
  Speaker Verification}, in: ICASSP 2022-2022 IEEE International Conference on
  Acoustics, Speech and Signal Processing (ICASSP), IEEE, 2022, pp. 6727--6731.

\bibitem{Mingote2018}
V.~Mingote, A.~Miguel, A.~Ortega, E.~Lleida,
  \href{http://dx.doi.org/10.21437/IberSPEECH.2018-1}{{Differentiable
  Supervector Extraction for Encoding Speaker and Phrase Information in Text
  Dependent Speaker Verification}}, Proceedings of IberSPEECH 2018 (2018)
  1--5\href {http://dx.doi.org/10.21437/IberSPEECH.2018-1}
  {\path{doi:10.21437/IberSPEECH.2018-1}}.
\newline\urlprefix\url{http://dx.doi.org/10.21437/IberSPEECH.2018-1}

\bibitem{mingote2019supervector}
V.~Mingote, A.~Miguel, A.~Ortega, E.~Lleida, {Supervector Extraction for
  Encoding Speaker and Phrase Information with Neural Networks for
  Text-Dependent Speaker Verification}, Applied Sciences 9~(16) (2019) 3295.

\bibitem{mingote2020optimization}
V.~Mingote, A.~Miguel, A.~Ortega, E.~Lleida, {Optimization of the area under
  the ROC curve using neural network supervectors for text-dependent speaker
  verification}, Computer Speech \& Language 63 (2020) 101078.

\bibitem{reynolds1995robust}
D.~A. Reynolds, R.~C. Rose, et~al., {Robust text-independent speaker
  identification using Gaussian mixture speaker models}, IEEE transactions on
  speech and audio processing 3~(1) (1995) 72--83.

\bibitem{reynolds2000speaker}
D.~A. Reynolds, T.~F. Quatieri, R.~B. Dunn, {Speaker verification using adapted
  Gaussian mixture models}, Digital signal processing 10~(1-3) (2000) 19--41.

\bibitem{rabiner1989tutorial}
L.~R. Rabiner, {A tutorial on hidden Markov models and selected applications in
  speech recognition}, Proceedings of the IEEE 77~(2) (1989) 257--286.

\bibitem{Mingote2021icassp}
V.~Mingote, A.~Miguel, A.~Ortega, E.~Lleida, {Memory Layers with Multi-Head
  Attention Mechanisms for Text-Dependent Speaker Verification}, in: ICASSP
  2021 - 2021 IEEE International Conference on Acoustics, Speech and Signal
  Processing (ICASSP), 2021, pp. 6154--6158.
\newblock \href {http://dx.doi.org/10.1109/ICASSP39728.2021.9414859}
  {\path{doi:10.1109/ICASSP39728.2021.9414859}}.

\bibitem{lample2019large}
G.~Lample, A.~Sablayrolles, M.~Ranzato, L.~Denoyer, H.~J{\'e}gou, Large memory
  layers with product keys, in: Advances in Neural Information Processing
  Systems, 2019, pp. 8546--8557.

\bibitem{cai2018novel}
W.~Cai, Z.~Cai, X.~Zhang, X.~Wang, M.~Li, A novel learnable dictionary encoding
  layer for end-to-end language identification, in: 2018 ICASSP, pp.
  5189--5193.

\bibitem{Hinton2015DistillingNetworkb}
G.~Hinton, O.~Vinyals, J.~Dean,
  \href{http://arxiv.org/abs/1503.02531}{{Distilling the Knowledge in a Neural
  Network}}, NIPS 2014 Deep Learning Workshop (2015) 1--9\href
  {http://dx.doi.org/10.1063/1.4931082} {\path{doi:10.1063/1.4931082}}.
\newline\urlprefix\url{http://arxiv.org/abs/1503.02531}

\bibitem{zhong2020random}
Z.~Zhong, L.~Zheng, G.~Kang, S.~Li, Y.~Yang, Random erasing data augmentation,
  in: Proceedings of the AAAI Conference on Artificial Intelligence, Vol.~34,
  2020, pp. 13001--13008.

\bibitem{mingote2020bdk}
V.~{Mingote}, A.~{Miguel}, D.~{Ribas}, A.~{Ortega}, E.~{Lleida}, {Knowledge
  Distillation and Random Erasing Data Augmentation for Text-Dependent Speaker
  Verification}, in: 2020 IEEE International Conference on Acoustics, Speech
  and Signal Processing (ICASSP), IEEE, 2020, pp. 6824--6828.

\bibitem{vinals2019phonetically}
I.~Vi{\~n}als, D.~Ribas, V.~Mingote, J.~Llombart, P.~Gimeno, A.~Miguel,
  A.~Ortega, E.~Lleida, {Phonetically-Aware Embeddings, Wide Residual Networks
  with Time-Delay Neural Networks and Self Attention Models for the 2018 NIST
  Speaker Recognition Evaluation}, Proc. Interspeech 2019 (2019) 4310--4314.

\bibitem{blundell2015weight}
C.~Blundell, J.~Cornebise, K.~Kavukcuoglu, D.~Wierstra, Weight uncertainty in
  neural network, in: International Conference on Machine Learning, PMLR, 2015,
  pp. 1613--1622.

\bibitem{Korattikara2015BayesianKnowledge}
A.~Korattikara, V.~Rathod, K.~Murphy, M.~Welling, {Bayesian Dark Knowledge},
  arXiv (2015) 1--9\href {http://dx.doi.org/10.1017/CBO9781107415324.004}
  {\path{doi:10.1017/CBO9781107415324.004}}.

\bibitem{shen2018feature}
P.~Shen, X.~Lu, S.~Li, H.~Kawai, {Feature Representation of Short Utterances
  Based on Knowledge Distillation for Spoken Language Identification.}, in:
  Interspeech, 2018, pp. 1813--1817.

\bibitem{shen2019interactive}
P.~Shen, X.~Lu, S.~Li, H.~Kawai, {Interactive Learning of Teacher-student Model
  for Short Utterance Spoken Language Identification}, in: Proc. ICASSP, IEEE,
  2019, pp. 5981--5985.

\bibitem{he2016deep}
K.~He, X.~Zhang, S.~Ren, J.~Sun, Deep residual learning for image recognition,
  in: Proceedings of the IEEE conference on computer vision and pattern
  recognition, 2016, pp. 770--778.

\bibitem{Larcher2014Text-dependentRSR2015}
A.~Larcher, K.~A. Lee, B.~Ma, H.~Li, {Text-dependent speaker verification:
  Classifiers, databases and RSR2015}, Speech Communication 60 (2014) 56--77.

\bibitem{das2018compensating}
R.~K. Das, M.~Madhavi, H.~Li, Compensating utterance information in fixed
  phrase speaker verification, in: 2018 Asia-Pacific Signal and Information
  Processing Association Annual Summit and Conference (APSIPA ASC), IEEE, 2018,
  pp. 1708--1712.

\bibitem{liu2019unified}
T.~Liu, M.~C. Madhavi, R.~K. Das, H.~Li, {A Unified Framework for Speaker and
  Utterance Verification.}, Proc. Interspeech 2019 (2019) 4320--4324.

\bibitem{liu2020speaker}
T.~Liu, R.~K. Das, M.~Madhavi, S.~Shen, H.~Li, {Speaker-Utterance Dual
  Attention for Speaker and Utterance Verification}, Proc. Interspeech 2020
  (2020) 4293--4297.

\bibitem{deepmine2019asru}
H.~Zeinali, L.~Burget, J.~Cernocky, {A Multi Purpose and Large Scale Speech
  Corpus in {Persian and English} for Speaker and Speech Recognition: the
  {DeepMine} Database}, in: Proc. ASRU 2019.

\bibitem{sdsvc2020plan}
H.~Zeinali, K.~A. Lee, J.~Alam, L.~Burget, {Short-duration Speaker Verification
  ({SdSV}) Challenge 2020: the Challenge Evaluation Plan.}, Tech. rep., arXiv
  preprint arXiv:1912.06311.

\bibitem{panayotov2015librispeech}
V.~Panayotov, G.~Chen, D.~Povey, S.~Khudanpur, {Librispeech: An {ASR} corpus
  based on public domain audio books}, in: 2015 ICASSP, pp. 5206--5210.

\bibitem{lozano2020but}
A.~Lozano-Diez, A.~Silnova, B.~Pulugundla, J.~Rohdin, K.~Vesel{\`y}, L.~Burget,
  O.~Plchot, O.~Glembek, O.~Novotn{\`y}, P.~Matejka, {BUT Text-Dependent
  Speaker Verification System for SdSV Challenge 2020.}, in: INTERSPEECH, 2020,
  pp. 761--765.

\bibitem{chen2020improving}
Z.~Chen, Y.~Lin, {Improving X-Vector and PLDA for Text-Dependent Speaker
  Verification.}, in: INTERSPEECH, 2020, pp. 726--730.

\bibitem{nagrani2017voxceleb}
A.~Nagrani, J.~S. Chung, A.~Zisserman, {{VoxCeleb}: A Large-Scale Speaker
  Identification Dataset}, in: Proc. Interspeech 2017, pp. 2616--2620.

\bibitem{chung2018voxceleb2}
J.~S. Chung, A.~Nagrani, A.~Zisserman, {{VoxCeleb2}: Deep Speaker Recognition},
  in: Proc. Interspeech 2018, 2018, pp. 1086--1090.

\bibitem{nist2008}
\href{https://www.nist.gov/sites/
  default/files/documents/2017/09/26/sre08$\_$evalplan$\_$release4.pdf}{{The
  NIST Year 2008 Speaker Recognition Evaluation Plan}} (2008).
\newline\urlprefix\url{https://www.nist.gov/sites/
  default/files/documents/2017/09/26/sre08$\_$evalplan$\_$release4.pdf}

\bibitem{nist2010}
\href{https://www.nist.gov/sites/
  default/files/documents/itl/iad/mig/NIST$\_$SRE10$\_$evalplan-r6.pdf.}{{The
  NIST Year 2010 Speaker Recognition Evaluation Plan}} (2010).
\newline\urlprefix\url{https://www.nist.gov/sites/
  default/files/documents/itl/iad/mig/NIST$\_$SRE10$\_$evalplan-r6.pdf.}

\bibitem{desplanques2020ecapa}
B.~Desplanques, J.~Thienpondt, K.~Demuynck, {ECAPA-TDNN: Emphasized Channel
  Attention, Propagation and Aggregation in TDNN Based Speaker Verification},
  Proc. Interspeech 2020 (2020) 3830--3834.

\end{thebibliography}

{\includegraphics[width=1in,height=1.25in,keepaspectratio]{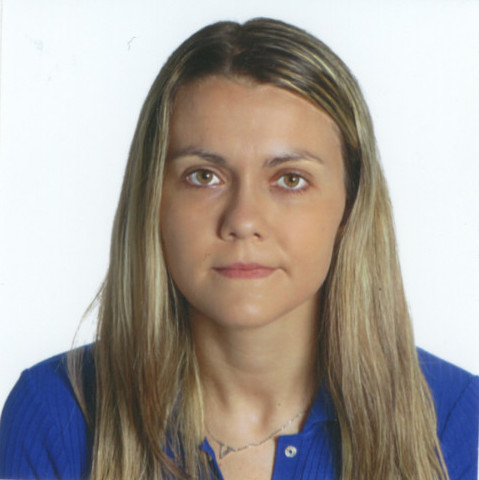}}
\textbf{Victoria Mingote} received the Bachelor’s and Master’s degree in Telecommunication Engineering from the University of Zaragoza, Spain, in 2014 and 2016, respectively. After that, she joined ViVoLab research group as a Ph.D student and received her Ph.D. degree in 2022 from the University of Zaragoza. She is currently a post-doctoral researcher in the ViVoLab research group. She has achieved several publications of her work in different international journals and conference proceedings. Her research interests expands through the areas of signal processing, machine learning, multimodal verification and identification (voice and face), and language identification.

{\includegraphics[width=1in,height=1.25in,clip,keepaspectratio]{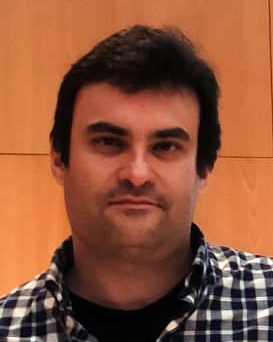}}
\textbf{Antonio Miguel} was born in Zaragoza, Spain. He received the M.Sc. degree in telecommunication engineering and the Ph.D. degree from the University of Zaragoza (UZ), Zaragoza, Spain, in 2001 and 2008, respectively. From 2000 to 2006, he was with the Communication Technologies Group, Department of Electronic Engineering and Communications, UZ, under a research grant. Since 2006, he has been an Associate Professor in the Department of Electronic Engineering and Communications, UZ. His current research interests include acoustic modeling for speech and speaker recognition.

{\includegraphics[width=1in,height=1.25in,clip,keepaspectratio]{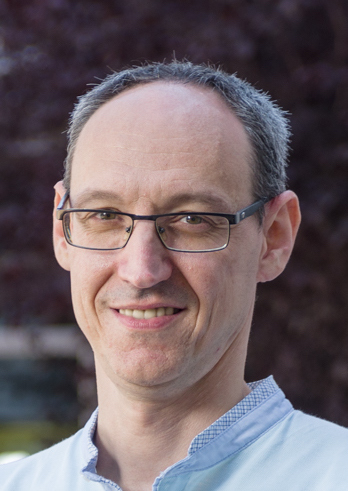}}
\textbf{Alfonso Ortega} received the Telecommunication Engineering and the Ph.D. degrees from the University of Zaragoza, Zaragoza, Spain, in 2000 and 2005, respectively. He is Associate Director of the Aragon Institute for Engineering Research (I3A), University of Zaragoza. In 2006, he was visiting scholar in the Center for Robust Speech Systems, University of Texas at Dallas, USA. He has participated in more than 50 research projects funded by national or international public institutions and more than 30 research projects for several companies. He is the author of more than 100 papers published in international journals or conference proceedings and several international patents. He is currently Associate Professor in the Department of Electronic Engineering and Communications, University of Zaragoza. His research interests include speech processing, analysis and modeling, automatic speaker verification, and automatic speech recognition. His Ph.D. thesis, advised by Dr. E. Lleida, received the Ph.D. Extraordinary Award and the Telefonica Chair Award to the best technological Ph.D.

{\includegraphics[width=1in,height=1.25in,clip,keepaspectratio]{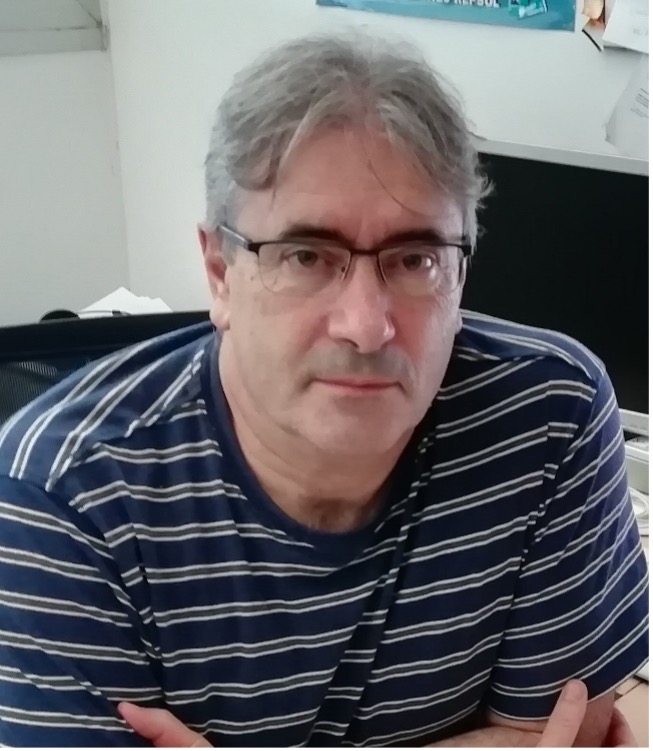}}
\textbf{Eduardo Lleida} received the M.Sc. degree in telecommunication engineering and the Ph.D. degree in signal processing from the Universitat Politecnica de Catalunya (UPC), Barcelona, Spain, in 1985 and 1990, respectively. From 1986 to 1988, he was involved in his doctoral work in the Department of Signal Theory and Communications, UPC. From 1989 to 1990, he was an Assistant Professor, and from 1991 to 1993, he was an Associated Professor in the Department of Signal Theory and Communications, UPC. From February 1995 to January 1996, he was with AT\&T Bell Laboratories, Murray Hill, NJ, USA, as a Consultant in speech recognition. He is currently a full Professor of signal theory and communications in the Department of Electronic Engineering and Communications, University of Zaragoza, Zaragoza, Spain, and a member of the Aragon Institute for Engineering Research, where he is heading the ViVoLab research group in speech technologies. He has been the doctoral advisor of 12 doctoral students. He has managed more than 50 speech-related projects, being inventor in seven worldwide patents, and coauthored more than 200 technical papers in the field of speech, speaker and language recognition, speech enhancement and recognition in adverse acoustic environments, acoustic modeling, confidence measures, and spoken dialogue systems.



\end{document}